\documentclass[Crown,sagev,times,doublespace]{sagej_dk}

\usepackage{moreverb,url}

\usepackage[colorlinks,bookmarksopen,bookmarksnumbered,citecolor=red,urlcolor=red]{hyperref}

\newcommand\BibTeX{{\rmfamily B\kern-.05em \textsc{i\kern-.025em b}\kern-.08em
T\kern-.1667em\lower.7ex\hbox{E}\kern-.125emX}}

\begin{document}

%\runninghead{Karlen}
\runninghead{}

\title{\\
Characterizing the spread of CoViD-19}

\date{July 13, 2020}

\author{Dean Karlen\affilnum{1,2}}
\affiliation{\affilnum{1}University of Victoria\\
\affilnum{2}TRIUMF}

\corrauth{Dean Karlen,
Department of Physics and Astronomy,
University of Victoria,
Victoria, BC, Canada, V8W 2Y2.}

\email{karlen@uvic.ca}

\begin{abstract}
Since the beginning of the epidemic,
daily reports of CoViD-19 cases, hospitalizations, and deaths
from around the world have been publicly available.
This paper describes methods to characterize
broad features of the spread of
the disease, with relatively long periods of
constant transmission rates,
using a new population modeling framework based on discrete-time difference equations.
Comparative parameters are chosen for their weak
dependence on model assumptions.
Approaches for their point and interval estimation, accounting
for additional sources of variance in the case data,
are presented.
These methods provide a basis to quantitatively assess the impact
of changes to social distancing policies using publicly available data.
As examples,
data from Ontario and German states are analyzed using this framework.
German case data show a small increase in transmission rates following the relaxation of lock-down
rules on May 6, 2020.
By combining case and death data from Germany, the mean and standard deviation of the time 
from infection to death are estimated.
\end{abstract}


\maketitle

% --------------------
\section{Introduction}
% --------------------

The CoViD-19 epidemic gained world-wide attention in March 2020
as the number of cases began to rise rapidly.
For most people, this was their first experience with an epidemic and
public health measures enacted to reduce social contact.
While these measures caused major disruptions in daily life, they
eventually reduced the rate of growth in case numbers.
Several regions saw localized outbreaks, particularly in long term care facilities and
meat processing facilities.
After a few months, social distancing regulations began to be relaxed, with the
intention that viral spread would kept at a manageable level.

While this describes the general experience in many western nations,
there are significant differences seen within some nations,
in their timelines and growth rates.
This paper presents methods to characterize differences
in the spread of CoViD-19 that can be deduced from
publicly available data.
This information can be useful for assessing the current situation
and to identify successful strategies to reduce transmission.

This paper introduces a new general-purpose
modelling framework developed to study the spread of CoViD-19.
A simple model, chosen to interpret CoViD-19 daily data, is described and
its properties are investigated.
Comparative statistics, chosen to have weak dependence on model assumptions,
are defined.
Data analysis approaches to estimate model parameters and their uncertainties are
described, along with studies of simulated data.

% --------------------
\section{Modeling framework}
% --------------------

%\lipsum[2-5] % Dummy text. Erase before write
%\citet{Hardaker2004} % Example of citation. Erase before use

The  python Population Modeller (pyPM\cite{pyPM}), is a general framework for building models of
viral spread using discrete-time difference equations.
A pyPM model is an object that consists of a set of population objects connected by an ordered list of 
directional connector objects.
Parameter objects are used to manage the various adjustable parameters necessary to define a specific model.
The object oriented design separates the task of model design from numerical implementation
which reduces the risk of implementation errors and simplifies the process of model redesign.
The model object, containing the model design and its parameters, can be saved in
small files, allowing for a multitude of models to be cataloged.

Two main reasons favor the use of discrete-time difference equations for this study, over the
traditional ordinary differential equation approach.
Firstly, with discrete-time difference equations, it is straightforward to implement arbitrary time
delay distributions, such as normal distributions with mean and standard deviations specified as
parameters.
Secondly, the purpose of the modeller is to interpret publicly available data reported
as daily counts, and with a step size of one day,
daily expectations or simulated daily counts are computed directly.
The daily values are saved as a time series in the population objects, and referred to as ``histories''.

The basic connector, called a propagator, transfers a fraction of incoming members from one population
(referred to as the ``from-population'')
to another (referred to as the ``to-population'') according to a delay distribution.
The delayed transfer is accomplished by having each population maintain a list of incoming contributions for each day in the future.
A ``splitter'' connector divides the incoming ``from-population'' members to two or more ``to-populations''.
A ``multiplier'' connector produces a number of new members in the ``to-population'', based on the sizes of the ``from-populations''
and forms the basis of the infection cycle.
``Subtractor'' and ``Adder'' connectors subtract from or add to populations without delay.

Prior to taking a time step, the connectors are processed in order, with each
calculating the future contributions to its ``to-populations'', given the number of ``from-population'' members arriving
at the next time step.
After all calculations are completed, the time step is taken, by having each population extend its history with a new value,
consisting of the previous day's number and the future contribution for that day.

As an illustration, consider a model in which a fraction $f_u$ of the symptomatic population $S$ are admitted
into ICU, population $U$, and those are split into two populations, a fraction $f_v$ are treated with a ventilator $V$, and the remaining are released without ventilation, $W$.
If the current time step is $t$, the expected incoming population for $S$ is $E[\Delta S_{t+1}]$. 
Future expected contributions to population $U$ arising from that incoming population to S are:
\begin{align*}
    E[\Delta U_{t+1+j}] &= f_u\, E[\Delta S_{t+1}] \, \beta_{uj} \\
    \beta_{u0} &= \int_{-\infty}^{\frac{1}{2}} g_u(t)\, dt \\
    \beta_{uj} &= \int_{j-\frac{1}{2}}^{j+\frac{1}{2}} g_u(t)\, dt \mathrm{\ \ \ for\ \ } j>0
\end{align*}
where $g_u$ is the distribution that defines the delay in the symptomatic population transferring to the ICU population.
Following that calculation, the future contributions to the ventilated and non-ventilated populations, $V$ and $N$, are calculated in
a similar fashion,
\begin{align*}
    E[\Delta V_{t+1+j}] &= f_v\, E[\Delta U_{t+1}] \, \beta_{vj} \\
    E[\Delta W_{t+1+j}] &= (1-f_v)\, E[\Delta U_{t+1}] \, \beta_{nj} \\
\end{align*}

Alternatively, instead of calculating expectation values,
the calculations can be performed as a stochastic simulation, where population sizes are defined by integers, for example,
\begin{align*}
    b &= B(\Delta S_{t+1},f_u) \\
    \Delta U_{t+1+j} &=  M(b,\beta_{uj}) \\
\end{align*}
where $B$ returns a binomial variate and $M$ returns multinomial variates.
The stochastic simulation of the infection cycle would use a Poisson variate if infections were described by independent
events.
To account for grouping of infections, negative binomial variates are used,
parameterized by the mean $\mu$ and $p_{nb} \in (0,1)$, such that the variance is $\mu/p_{nb}$.
The stochastic calculations are useful for checking for bias and for evaluating standard deviations of estimators.

% --------------------
\section{Model for characterizing the spread of CoViD-19}
% --------------------

The pyPM reference model 2.3 is a simple model developed to characterize the spread of CoViD-19.
It includes an infection cycle connected to populations that correspond to publicly available
data for cases, deaths, and hospitalization, including ICU and ventilator use.

% --------------------
\subsection{Infection cycle model}
% --------------------

Results in this paper use pyPM reference model 2.3 as the nominal model, and its
infection cycle is defined by three connectors.
Firstly, for each day, the expected number of new infections is calculated as:
$$
E[\Delta I_{t+1}] = \alpha \frac{E[S_t]}{E[N_t]} E[C_t]
$$
where $\alpha$ is the transmission rate, $S$ the susceptible population, $N$ the total population, and $C$ the
circulating contagious population.
In the initial stages of an epidemic, when almost the entire population is susceptible,
$\alpha$ represents the average number of people that a contagious person infects in one day.
Its value can be considered to be linearly proportional to the number of contacts and closeness of contact between individuals,
and therefore it is a parameter that is linearly proportional to social distancing measures.

The second connector propagates 
a fraction of the newly infected population to the circulating contagious population, with a delay distribution modeling
the so-called ``latent period''.
The third connector reduces the size of the
circulating contagious population using a propagator to a removed population and a subtractor is used to remove them.
The propagator uses a delay distribution to represent the ``contagious circulation period''
which arises from all ways that contagious individuals become unable
to infect others, including quarantine, hospitalization, natural recovery, or death.

In the initial stages and when the transmission rate and circulation time are constant,
the ``steady state'' solution to the time difference equations has the
expected contagious population being exponential in time,
$$E[C_{t+1}] = (1+\delta)E[C_{t}].$$
The parameter describing fractional daily growth, identified as $\delta$ in this paper, is
often referred to as $r$ in epidemiology literature.
To ensure that the initial state corresponds to a ``steady state'' solution, the model
uses a ``boot-strap'' approach.
A state with a very small expectation value for the contagious population size is allowed
to grow until the target contagious population for the initial state is reached.
While only the initial contagious population needs to be specified, all other
populations will be assigned non-zero expectation values at the initial state that
correspond to a ``steady state'' solution.

The infection cycle model is purposefully simplistic, being described by a homogeneous population.
This reduces the number of parameters and characterizes the epidemic spread for
those groups in the population that contribute the greatest numbers to
cases, hospitalization, or deaths.
While the pyPM framework includes ensembles for modelling heterogeneous populations,
there is little public data available to constrain the many additional parameters.
For this paper, we characterize the spread using a homogeneous model.

% --------------------
\subsection{Modeling case reporting, hospitalization, and deaths}
% --------------------

The pyPM reference model 2.3 connects the contagious population to
the symptomatic population and from symptomatic to
reported (positive test cases), icu admission, and non-icu admission populations.
It also has an independent propagator from contagious to recovery or death.
Each of these connectors is parametrized by a fraction and a normal delay distribution.

If testing captures a constant fraction of contagious individuals, then the 
expected cases will follow the contagious population trajectory with a time lag.
The case data can therefore be used to estimate $\delta$ during these periods
without reference to a particular model.
Hospitalization and death rates provide additional measures of the infection trajectory.
When these are seen to follow different growth curves, this may indicate differences
in transmission by age or other factors, since these data are
unequal samplings of the infected population.

% --------------------
\subsection{Model dependence on transmission rate estimators}\label{sec:bias_alpha}
% --------------------

It would be useful to characterize the phases of an epidemic with a transmission rate parameter
like $\alpha$ that in some sense is linearly proportional to social distancing.
By doing so, statements can be made about relative levels of social distancing observed in past phases
and levels required going forward.
Unfortunately, unlike for $\delta$,
estimating the transmission rate, $\alpha$, from case data
sensitively depends on
the latent and circulation delay distributions which are not well known.
The well known parameter to characterize the growth rate, 
the reproduction number $R_0$, also suffers from strong model 
dependence~\cite{Li2011,Wallinga2007}.

\begin{figure}[t]
    \centering
        \includegraphics[scale=.56]{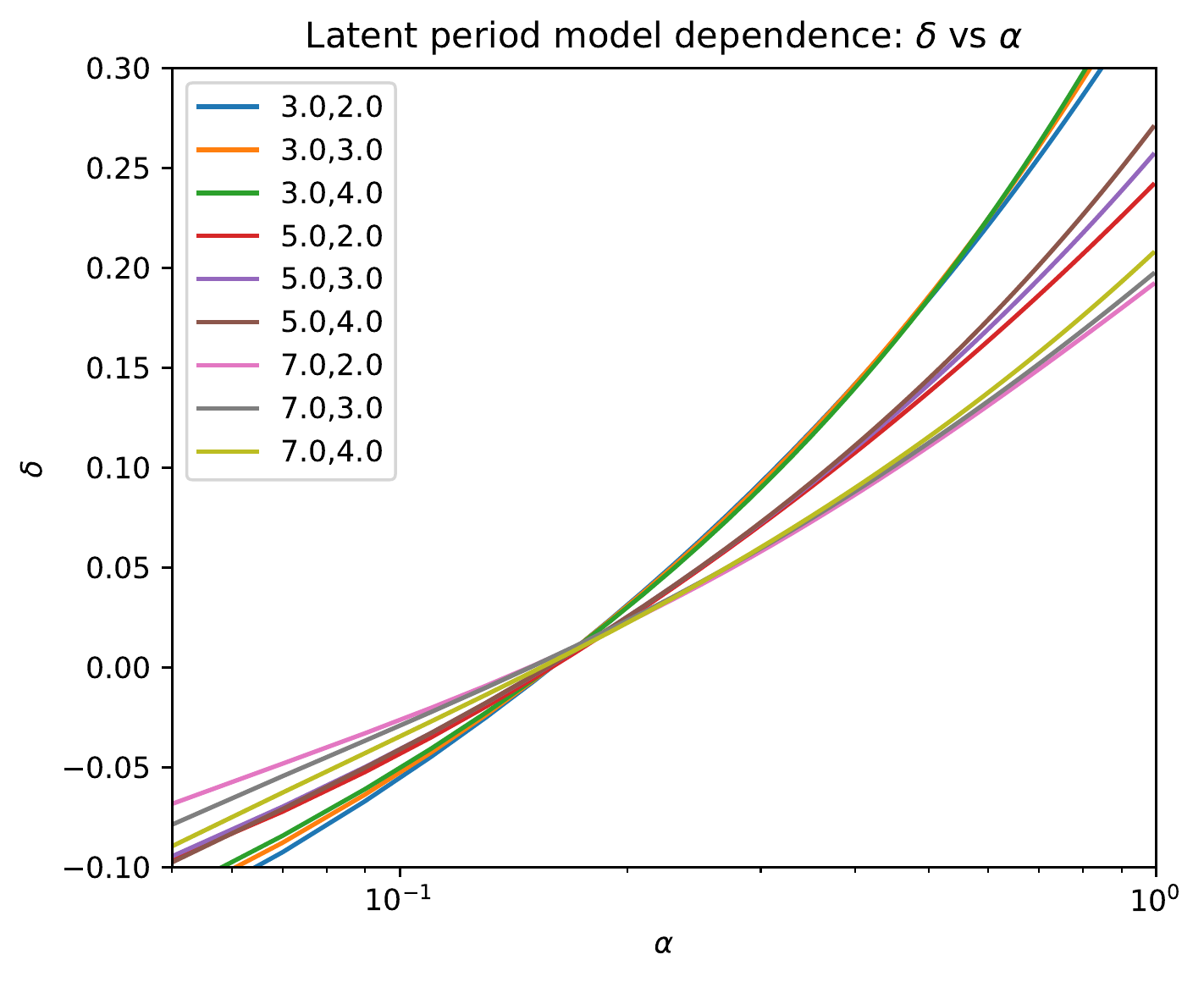}
        \includegraphics[scale=.56]{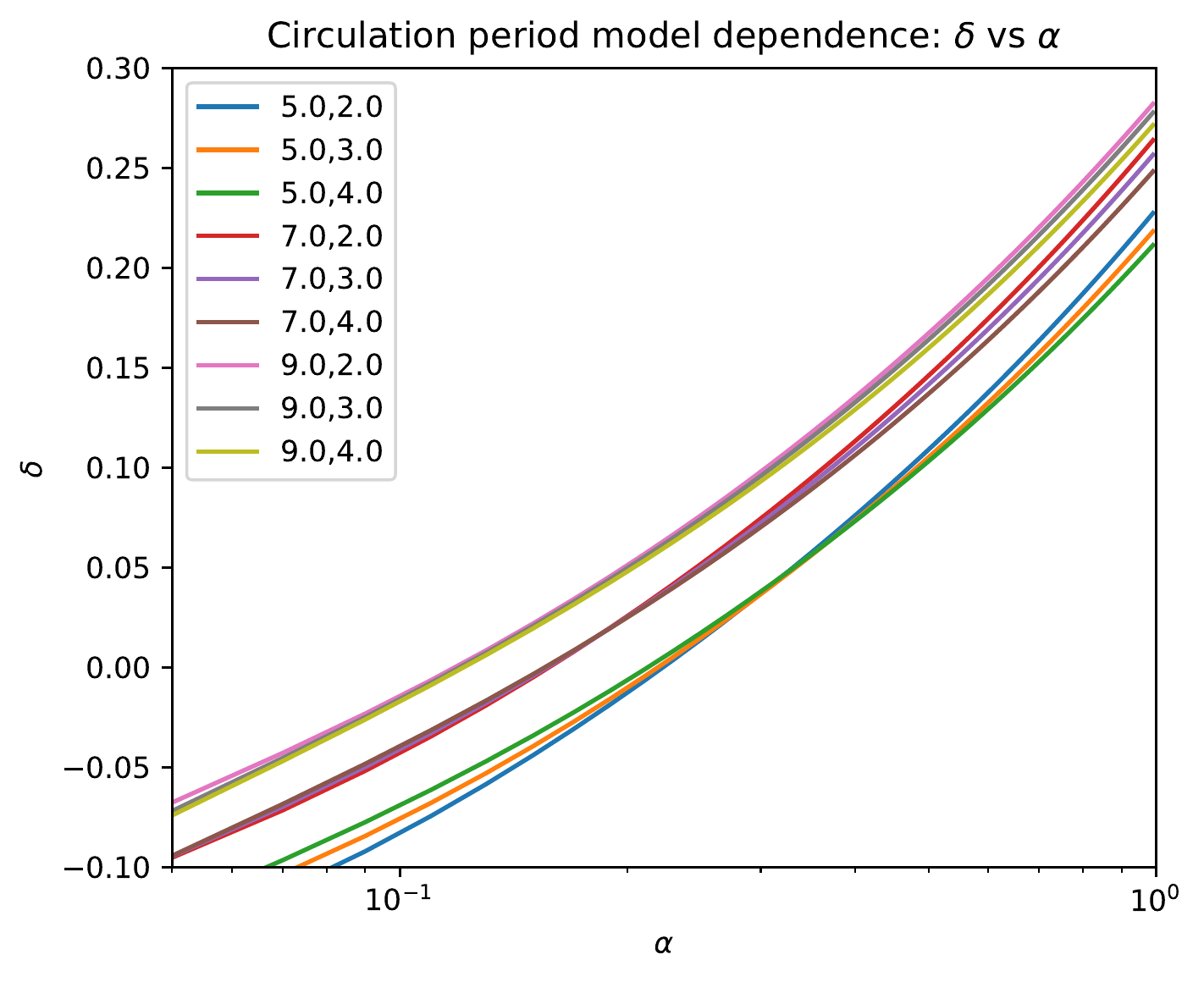}
    \caption{The relation between the exponential growth rate parameter $\delta$ and
    the transmission rate parameter $\alpha$ is shown for several choices
    for the latent period parameters (left) and circulation period parameters (right).
    The legend shows the means and standard deviations for the latent (left) and
    circulation (right) normal delay distributions (in days). The nominal model
    has $\ell_\mu=5$, $\ell_\sigma=3$, $c_\mu=7$, $c_\sigma=3$.}
    \label{fig:delta_vs_alpha}
\end{figure}

Figure~\ref{fig:delta_vs_alpha} shows how changing the delay distributions affects the
relation between the exponential growth and the transmission rate.
Increasing the mean circulation period increases $\delta$, since contagious
individuals have more time to infect more people.
Increasing the latent period reduces $\delta$ for a growth phase and
increases $\delta$ for a decline phase, since newly infected individuals take longer
to become contagious thereby reducing the feedback responsible for producing
the exponential growth or decline.

To further illustrate the sensitivity to the delay distribution parameters, 
consider the default conditions for pyPM reference model 2.3, which
describes a period of growth followed by a period of decline:

\begin{itemize}
    \item $C_0$ (initial contagious population): \texttt{cont\_0 = 10}
    \item $\alpha_0$ (initial transmission rate): \texttt{alpha\_0 = 0.4}
    \item $t_1$ (day \# for transmission rate change): \texttt{trans\_rate\_1\_time = 20}
    \item $\alpha_1$ (new transmission rate): \texttt{alpha\_0 = 0.1}
    \item $\ell_\mu$ (latent period mean): \texttt{cont\_delay\_mean = 5} (days)
    \item $\ell_\sigma$ (latent period standard deviation): \texttt{cont\_delay\_sigma = 3} (days)
    \item $c_\mu$ (circulation period mean): \texttt{removed\_delay\_mean = 7} (days)
    \item $c_\sigma$ (circulation period standard deviation): \texttt{removed\_delay\_sigma = 3} (days)
\end{itemize}

If case data produced according to this nominal model, is analyzed with a modified model, 
having different assumptions for the latent and circulation periods,
the estimators for $\alpha_0$ and $\alpha_1$ will be biased.
To estimate the bias, assume that estimators for the model independent
parameters, $\delta_0$ and $\delta_1$, are unbiased.
Given a choice for the delay parameters, the relation between $\hat\alpha$ and $\hat\delta$ can be found empirically
$$
\hat\alpha = \hat\alpha(\hat\delta \,|\, \ell_\mu, \ell_\sigma, c_\mu, c_\sigma),
$$
which would be the inverse of the functions shown in Fig.~\ref{fig:delta_vs_alpha}.
With sufficient statistics, the bias in the transmission rate estimators would be approximately
$$
b = E[\hat\alpha] - \alpha = \hat\alpha(E[\hat\delta] \,|\, \ell_\mu, \ell_\sigma, c_\mu, c_\sigma) - \alpha.
$$
Table~\ref{tab:alpha_bias} shows the expectation values and the
relative bias for the transmission rate estimators for reasonable alternative delay parameter values.
For many cases,
the bias is larger than typical standard deviations of the estimators (statistical uncertainty).
When alternative latent period parameters are used, the bias for the growth and decline estimators
have opposite sign.

\begin{table}[t]
\small\sf\centering
\caption{The size of transmission rate estimator biases for alternative delay parameter values
for an epidemic having a growth period followed by decline, as described in the text.
The last column shows that the ratio of the estimated transmission rates varies from 3 to 6 
depending on the delay parameters.}
\label{tab:alpha_bias}
\begin{tabular}{c c c c c c c c c} 
\toprule
 $\ell_\mu$ & $\ell_\sigma$ & $c_\mu$ & $c_\sigma$ & $E[\hat\alpha_0]$ & $\frac{b_0}{\alpha_0}$ (\%) & $E[\hat\alpha_1]$ & $\frac{b_1}{\alpha_1}$ (\%) & $E[\hat\alpha_0]/E[\hat\alpha_1]$\\
\midrule
3 & 2 & 7 & 3 & 0.333 & -17 & 0.114 & 14 & 2.9 \\
3 & 3 & 7 & 3 & 0.334 & -16 & 0.112 & 12 & 3.0 \\
3 & 4 & 7 & 3 & 0.338 & -15 & 0.110 & 10 & 3.1 \\
5 & 2 & 7 & 3 & 0.407 & 2 & 0.103 & 3 & 4.0 \\
5 & 3 & 7 & 3 & 0.400 & 0 & 0.100 & 0 & 4.0 \\
5 & 4 & 7 & 3 & 0.396 & -1 & 0.100 & 0 & 4.0 \\
7 & 2 & 7 & 3 & 0.497 & 24 & 0.079 & -21 & 6.3 \\
7 & 3 & 7 & 3 & 0.489 & 22 & 0.085 & -15 & 5.8 \\
7 & 4 & 7 & 3 & 0.477 & 19 & 0.092 & -8 & 5.2 \\
5 & 3 & 5 & 2 & 0.503 & 26 & 0.154 & 54 & 3.3 \\
5 & 3 & 5 & 3 & 0.516 & 29 & 0.148 & 48 & 3.5 \\
5 & 3 & 5 & 4 & 0.524 & 31 & 0.139 & 39 & 3.8 \\
5 & 3 & 7 & 2 & 0.390 & -2 & 0.102 & 2 & 3.8 \\
5 & 3 & 7 & 3 & 0.400 & 0 & 0.100 & 0 & 4.0 \\
5 & 3 & 7 & 4 & 0.411 & 3 & 0.098 & -2 & 4.2 \\
5 & 3 & 9 & 2 & 0.330 & -18 & 0.072 & -28 & 4.6 \\
5 & 3 & 9 & 3 & 0.336 & -16 & 0.074 & -26 & 4.5 \\
5 & 3 & 9 & 4 & 0.344 & -14 & 0.075 & -25 & 4.6 \\
\bottomrule
\end{tabular}
\end{table}

From this study, we find that transmission rates, or relative transmission rates, are not good
choices for characterizing growth or decline due to their sensitivity to poorly 
known model parameters.
Instead, in this paper we use the nominal model parameters to form estimators for $\alpha$ 
and convert those to
estimators for $\delta$.
As shown in section~\ref{sec:point}, the biases in estimators for $\delta$ are found to be small.

\newpage

% --------------------
\subsection{Model dependence on contagious population estimators}\label{sec:uc}
% --------------------

In addition to the rate of growth or decline, represented by $\delta$,
the size of the circulating contagious population is important to characterize the state of the epidemic.
This is not directly measured by case data and estimators are affected by many unknown factors, such as the fraction of
infected individuals who are tested.
While the absolute number may be poorly known, a relative indicator that has weaker model dependence
would be useful to indicate relative prevalence
between two regions or between two different periods within a region.
For this purpose, case data is analyzed using the nominal model, and the deduced contagious population
is scaled by the ratio of total cases to total infections.
The result, $UC$, for ``uncorrected circulating contagious population'', would need to be divided by the fraction of
infected individuals tested, to be an estimate for the absolute size of the contagious population.
The correction factor is not necessary to compare the relative prevalence in regions with similar testing policy.

\begin{figure}[t]
    \centering
        \includegraphics[scale=.475]{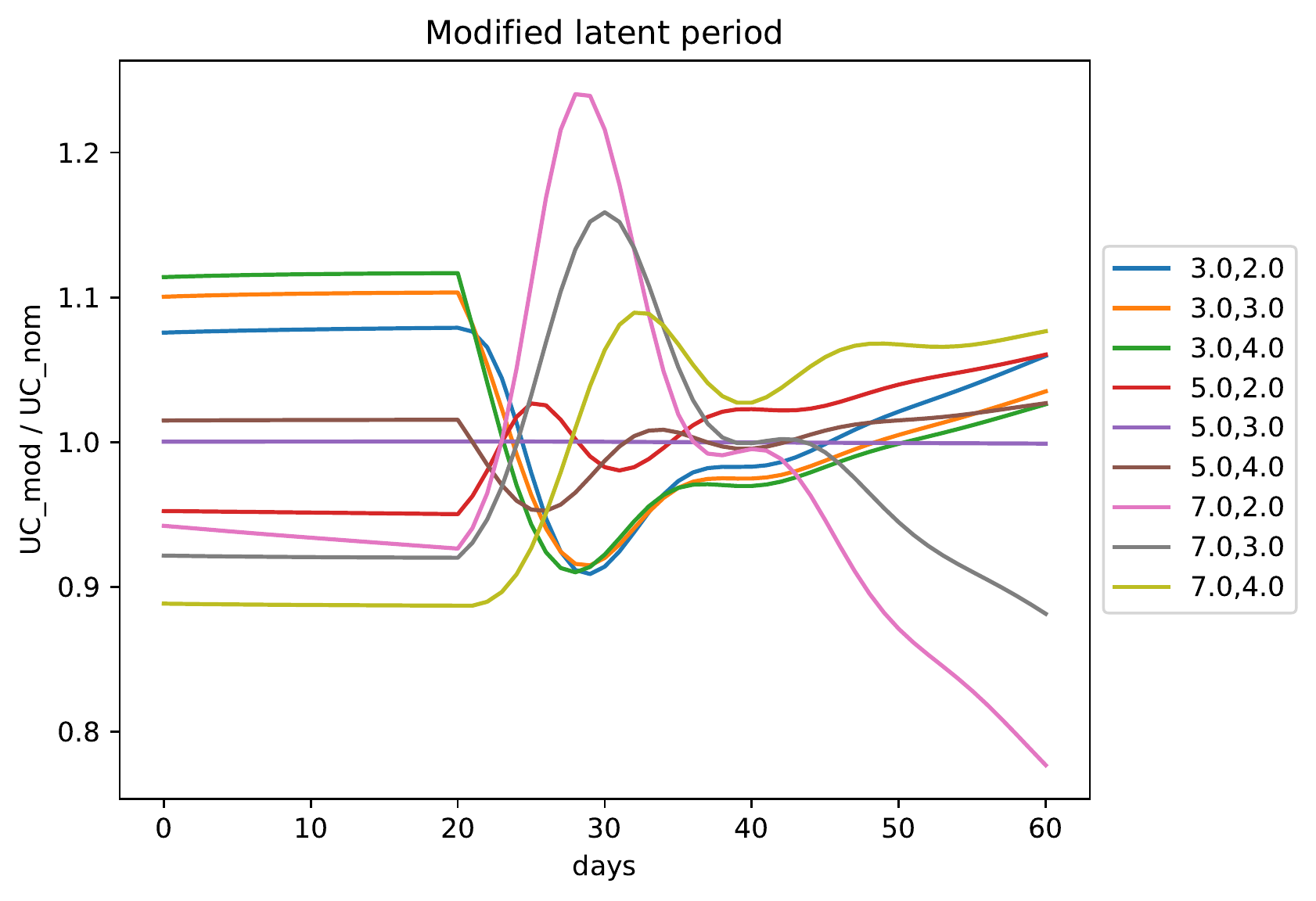}
        \includegraphics[scale=.475]{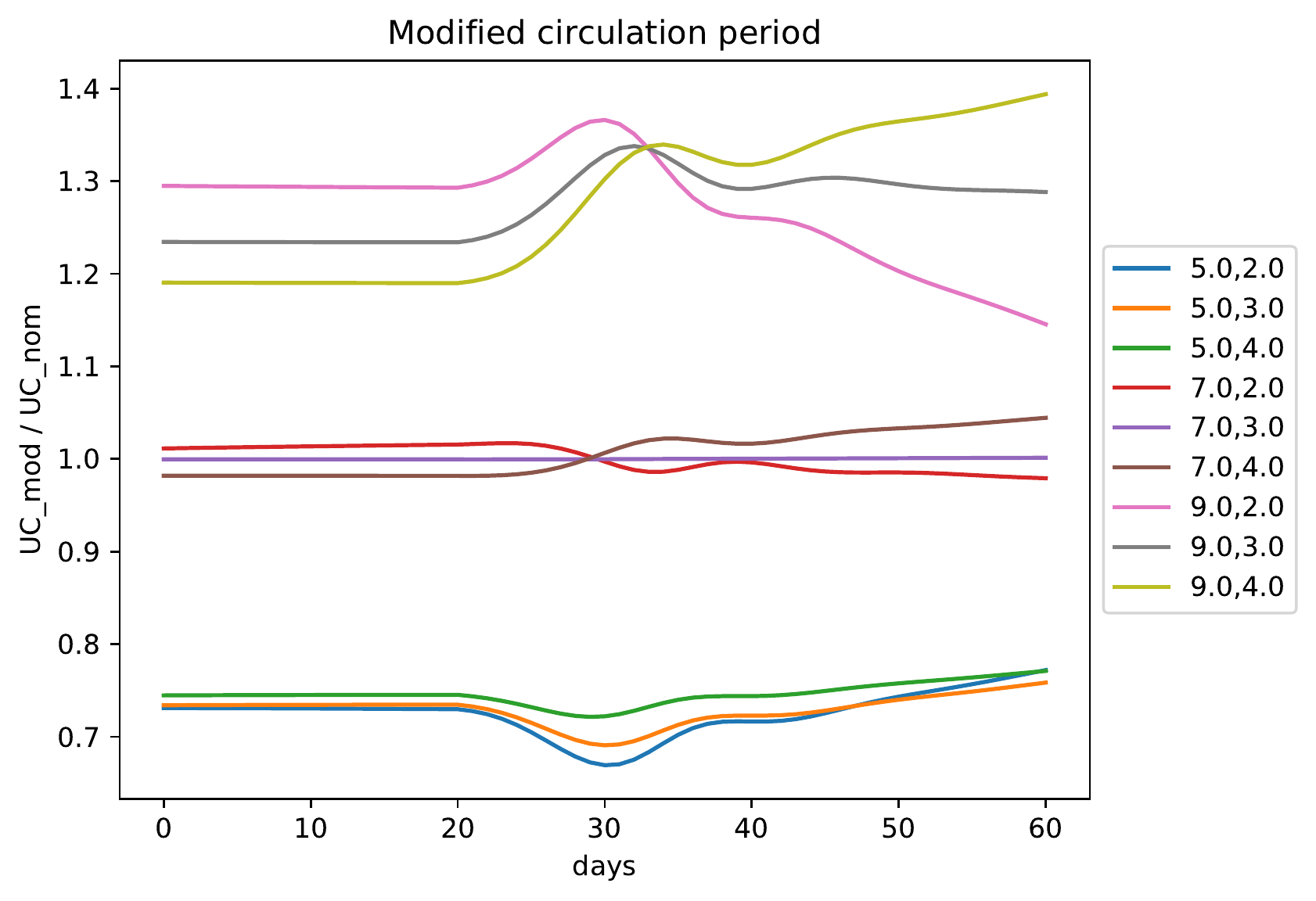}
    \caption{The expectation of the estimator for ``uncorrected circulating contagious population'' using a modified model compared to that
    using the nominal model. The adjusted latent period and circulation delay parameters ($\mu$ and $\sigma$) 
    are shown in the legend.}
    \label{fig:uc_bias}
\end{figure}

Just as for $\alpha$, the estimator for $UC$ depends on the latent and circulation delay parameters.
To illustrate the sensitivity, the same combinations for delay parameters are considered and the
ratio of $UC_\mathrm{mod}/UC_\mathrm{nom}$ is shown in Fig.~\ref{fig:uc_bias}
for the same conditions as described in section~\ref{sec:bias_alpha}.
Large deviations up to about 30\% are seen, but this effect is small compared to the
observed range of prevalence which spans several orders of magnitude.

% --------------------
\subsection{Modeling localized infection outbreaks}
% --------------------

If a large localized
infection outbreak occurs during a period where social distancing policy is being followed consistently,
the indicators for nominal infection growth/decline should be unaffected.
The rapid growth in count rates is not a result of changing social behaviour that would lead to a new
infection trajectory.
Instead, once the outbreak runs its course, 
the region would continue with the same growth/decline as before the outbreak.

To model this situation, 
a burst of infections is injected into the infected population, with the number and
date of burst optimized to match the case data.
During a period of increasing social distancing, a rapid rise in cases would be a clear signal for
an infection outbreak.
During a period when social distancing rules are being relaxed, an increase in cases could be a result of
the expected increase in transmission or due to a new outbreak.
In absence of other information that can distinguish these two hypotheses, it may be necessary to wait until
for the case data itself to identify whether or not the region is experiencing a new rate of growth.

% --------------------
\subsection{Modeling case reporting}\label{sec:case_reporting}
% --------------------

The daily number of new cases in a region is the publicly available indicator with the largest statistics 
and smallest time lag.
The pyPM reference model 2.3 models case reporting by propagating a fraction ($f_s=0.9$) 
of the newly contagious population to
the symptomatic population and a fraction ($f_r=0.8$) of those to the reported population.
The propagation time delays are described by normal distributions ($\mu_s=2$, $\sigma_s=1$ days) and
($\mu_r=3$, $\sigma_r=1$ days) respectively.
The size of the reported population is compared directly with publicly available case data
to characterize the spread of CoViD-19.
The uncertainties in the fractions contribute to systematic uncertainty in estimating the size of
the contagious populations, and the uncertainties in the delay parameters contribute to
uncertainty in identifying transition dates from the case data.

During the initial phase of the epidemic in western nations, in March 2020, 
testing policy and availability were changing significantly.
This may account for the fact that in many regions, including in Canada and the United States,
the number of cases did not follow an exponential growth trajectory in early March.
After that, the growth in number of cases are generally well described by models with 
relatively long periods of constant transmission and testing rates,
even though testing availability generally increased.
This suggests that revised testing policies enacted after March did not substantially change the fraction of
infected individuals getting tested.
Should a testing or reporting policy change cause a rapid rise in cases, this could be misinterpreted
as a reporting anomaly or a localized outbreak, but this will not cause
$\delta$ to be overestimated.
It is the
exponential growth of cases that determines $\delta$, not the absolute number of cases.

Estimating the growth and size of the contagious population from case data, and interval estimation in particular,
presents a challenge due to the large variance in the daily reported numbers.
Occasionally, a very large number of cases is reported on one day, as a backlog of cases clears some bottleneck in the
reporting process.
These anomalies are normally announced and they are modelled by an injection into the reported population.
Even ignoring those rare situations, daily case numbers have variance that far exceeds that expected
in a model with independent infected individuals being tested as they become symptomatic.
The pyPM model includes variation that arises from the reporting process itself, in which a variable fraction
of cases are held back and reported the next day. 
The fraction is drawn from a uniform distribution between $rn$ and 1, where $rn$ is the reporting noise parameter.
This produces a negative autocorrelation between one day and the next, which can be compared with actual data.
A second parameter, noise backlog, is provided to allow that only a fraction of the backlog being reported the next
day, to reduce the autocorrelation effect in the data.
By including this additional source of variance, the intervals for the growth parameter estimates can grow by a factor of 2 or more.

\begin{figure}[t]
    \centering
        \includegraphics[scale=.59]{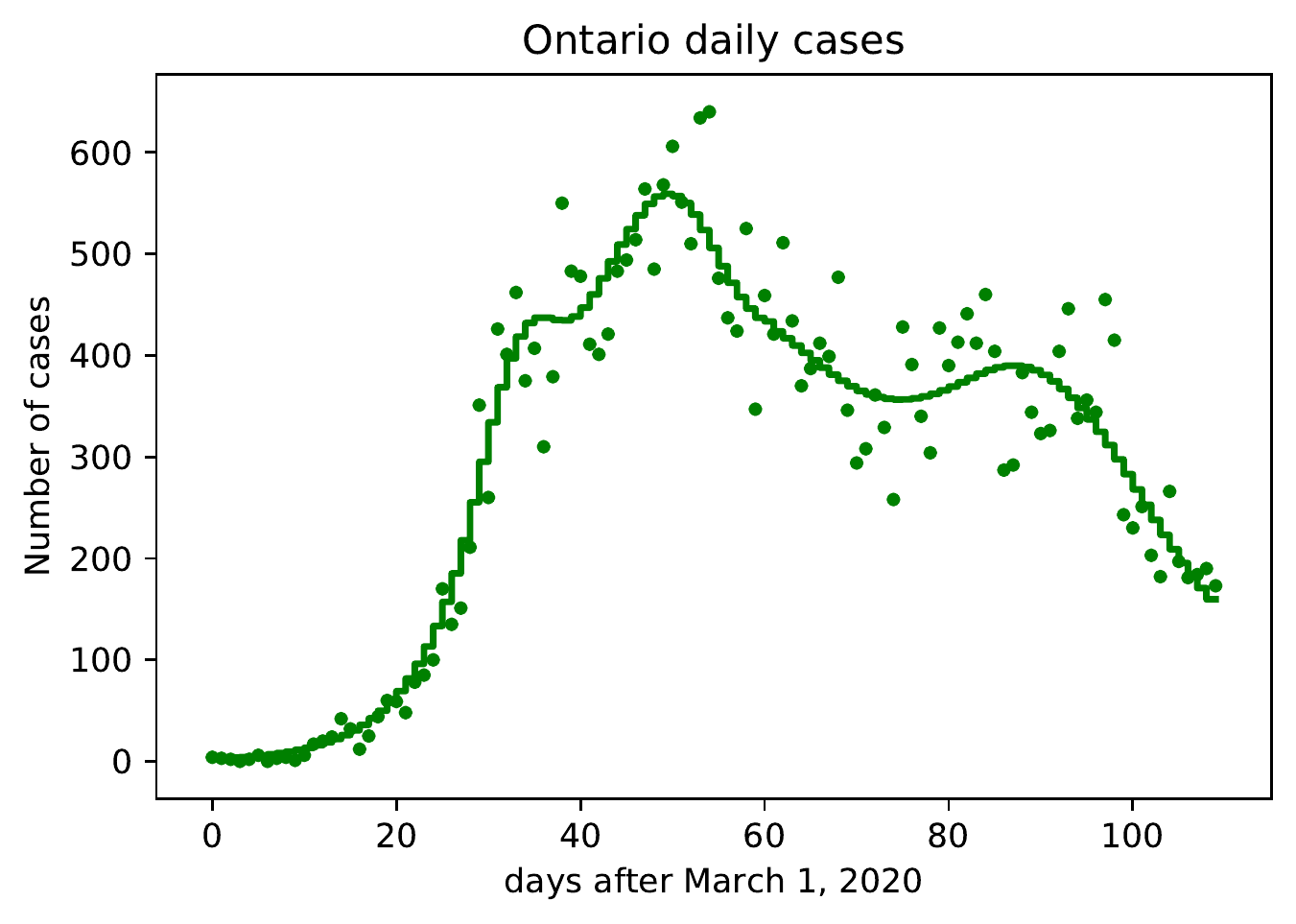}
        \includegraphics[scale=.59]{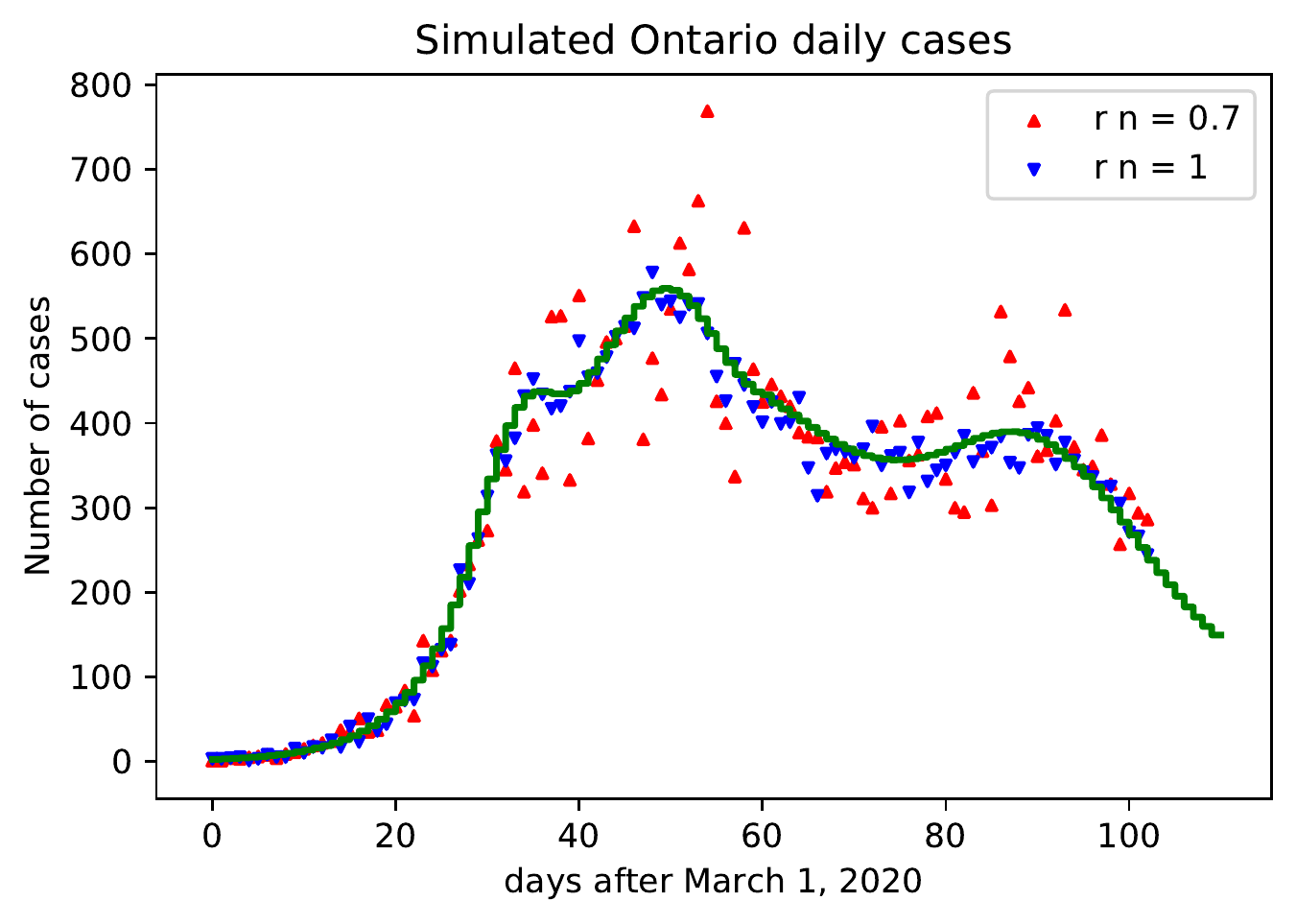}
    \caption{Left: Points show the daily cases reported from the province of Ontario and the curve shows the 
    expectations for each day from the model.
    Right: Simulations with no reporting noise ($rn$=1) and
    with reporting noise ($rn$=0.7) tuned to match the data goodness of fit statistic.
    The simulations use the negative binomial parameter for the infection cycle $p_{nb}$=0.1.
    To help the visual comparison, simulations were chosen that have total infections close to
    the expected value.}
    \label{fig:ont_cases}
\end{figure}

As an example, the daily cases from the province of Ontario is shown, along with the expectations from the model fit to that data
in Fig.~\ref{fig:ont_cases}.
The first 110 days are generally described by two transitions of transmission rate (on day 26 and 44) 
and one broad reporting anomaly centered at day 91. 
The day-to-day variation is larger than the model with no reporting noise.
In many jurisdictions, the effect of reporting noise is even larger than this.
To match the goodness of fit, calculated by assuming the daily data follow a Poisson distribution,
the reporting noise parameter is set to $0.7$.
Other provincial and state data typically require additional reporting noise
with this parameter set between 0 and 0.5.
To match the goodness of fit for the cumulative cases, the negative binomial parameter for the infection cycle $p_{nb}$
is set to 0.1.
These parameters were set by looking at the goodness of fit for 10 fitted simulation samples.

% --------------------
\subsection{Modeling hospitalization and death reporting}\label{sec:hosp_reporting}
% --------------------

The pyPM reference model 2.3 models hospitalization data by propagating a fraction 
of the newly symptomatic population to
icu and non-icu hospitalized populations, and a fraction of the icu admissions are
propagated to the ventilated population.
Each of these use by normal time delay distributions.
Deaths are modelled by propagating a fraction of the newly contagious population with
a normal time delay distribution.

% --------------------
\section{Point estimates using case data}\label{sec:point}
% --------------------

\begin{figure}[t]
    \centering
        \includegraphics[scale=.6]{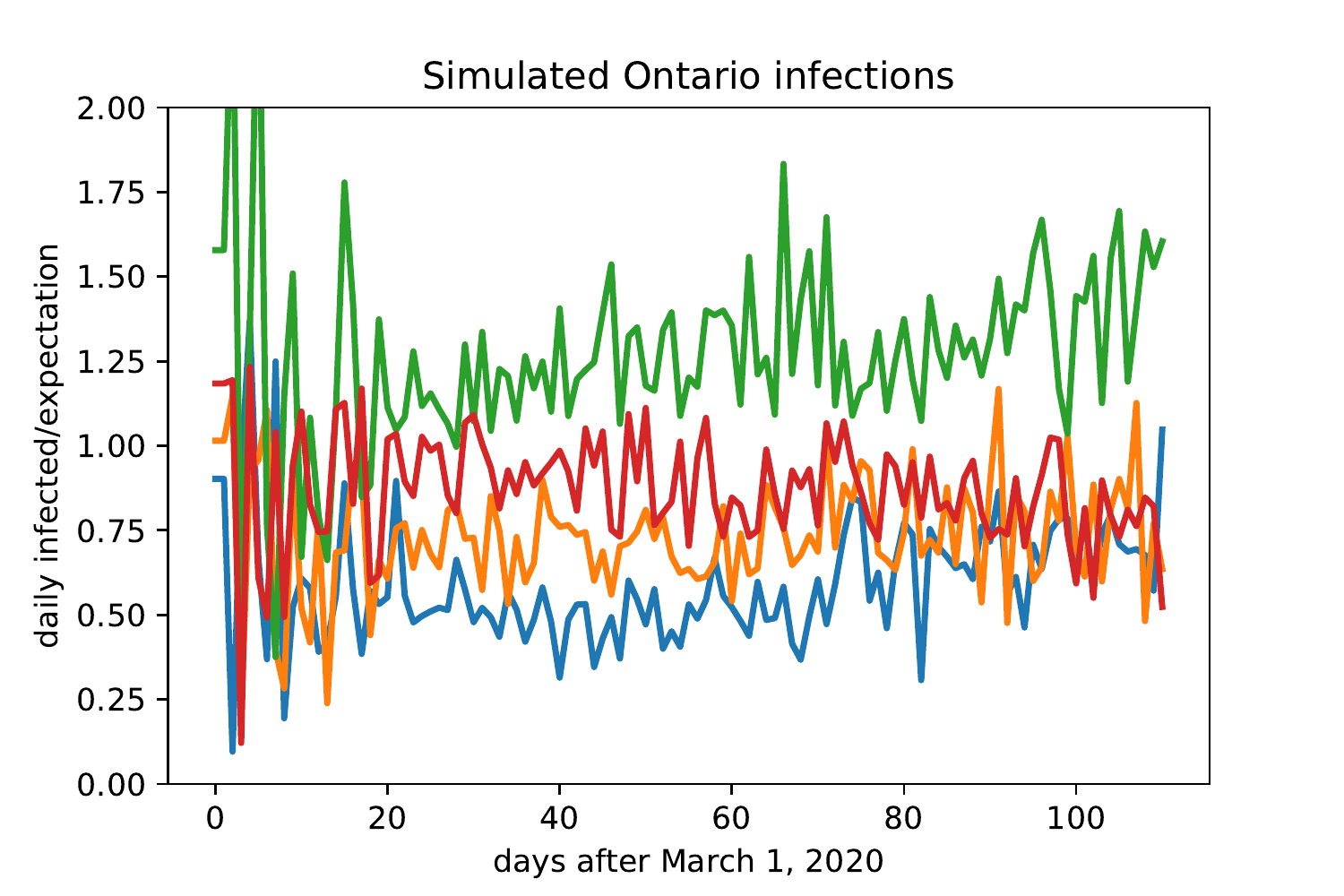}
    \caption{The ratio of daily new infections to expected new infections are shown for four simulations of the Ontario model to
    illustrate that daily new cases are not outcomes of independent random variables.
    The simulations use the negative binomial parameter for the infection cycle $p_{nb}$=0.1.}
    \label{fig:ont_infections}
\end{figure}

As indicated in section~\ref{sec:case_reporting}, the variance in the daily case counts exceeds that expected in a simple
model, and therefore an additional source of variation is included in the model to represent variation in daily reporting.
A more significant complication is that the daily cases do not represent outcomes of independent random variables.
Having a larger than expected number of infections for one day will generally result in larger than expected number of cases 
(and a larger than expected number of infections) in subsequent days.
This effect is illustrated in Fig.~\ref{fig:ont_infections}.

The standard approach of maximum likelihood estimation is difficult to apply given
the challenges in defining an appropriate likelihood function.
Instead, point estimates for model parameters are the combination that best reproduces the
cumulative case history by minimizing
the sum of the squares of the differences between the model expectations and data.
The following model parameters are estimated:

\begin{itemize}
    \item initial size of contagious population 
    \item transmission rates for each period (and end dates)
    \item sizes of infection outbreaks (and dates)
    \item size of reporting anomalies (and dates)
\end{itemize}

Biases and standard deviations of the estimators are found by fitting simulated samples.
As an example,
the distributions of point estimates for 1000 simulated samples using the Ontario model
are found to each be approximately normal
and only the final two ($\alpha_2$ and number in reporting anomaly) have strong correlation ($\rho\approx-0.6$).
The standard deviations of the transmission rate estimators are found to be approximately 5\% of the true values.
Table~\ref{tab:sim_ont_fits} shows that the biases are less than 1 standard deviation for all parameters.

\begin{table}[t]
\small\sf\centering
\caption{Descriptive statistics for point estimates for the 1000 simulated samples using the Ontario model.
The column labeled $\sigma_\mathrm{stat}$ shows the standard deviation of the point estimates.}
\label{tab:sim_ont_fits}
\begin{tabular}{l c c c c}
\toprule
    parameter & truth & mean & $\sigma_\mathrm{stat}$ & bias (in $\sigma_\mathrm{stat}$) \\
    \midrule
    \texttt{cont\_0} & 29.0 & 26.4 & 6.3 & -0.41 \\
    \texttt{alpha\_0} & 0.632 & 0.638 & 0.033 & 0.19 \\
    \texttt{alpha\_1} & 0.202 & 0.205 & 0.007 & 0.42 \\
    \texttt{alpha\_2} & 0.128 & 0.127 & 0.004 & -0.03 \\
    \texttt{anomaly\_1\_n} & 3727 & 3722 & 598 & -0.01 \\
\bottomrule
\end{tabular}
\end{table}

As discussed in section~\ref{sec:bias_alpha}, characterizing the growth in terms of $\delta$ instead of $\alpha$
has the benefit of being less model dependent.
This is illustrated in table~\ref{tab:ont_fits} which shows the point estimates for $\alpha$ and $\delta$
for different choices for the
parameters that define the latent and circulation periods in the infection cycle.
For this distribution of delay parameters,
the standard deviation of the $\alpha$ parameter estimates, indicated in the table as $\sigma_\mathrm{sys}$,
are significantly larger than standard deviations in the estimators, $\sigma_\mathrm{stat}$.
For $\delta$, the systematic and statistical standard deviations are nearly the same.
The systematic uncertainty is characterized by $\sigma_\mathrm{sys}$, provided the range of variation of the
delay parameters corresponds to about 1 standard deviation in the prior beliefs for these quantities.
The range of the delay parameters were chosen using the summary provided by
the Public Health Agency of Canada Modelling Group~\cite{PHAC}.

\begin{table}[t]
\small\sf\centering
\caption{Estimates for growth parameters for Ontario data from March 1 - June 17
under different latent and circulation period parameters.
For these fits, the transmission rate transition dates and the outbreak date were fixed.
Systematic uncertainty is characterized by the standard deviation of the parameter estimates for the distribution of
delay period parameters and statistical uncertainties are taken from Table~\ref{tab:sim_ont_fits}.}
\label{tab:ont_fits}
\begin{tabular}{c c c c c c c c c c} 
\toprule
 $\ell_\mu$ & $\ell_\sigma$ & $c_\mu$ & $c_\sigma$ & $\hat\alpha_0$ & $\hat\alpha_1$ & $\hat\alpha_2$ & 
 $\hat\delta_0$ & $\hat\delta_1$ & $\hat\delta_2$\\
\midrule
3 & 2 & 7 & 3 & 0.514 & 0.197 & 0.135 & 0.188 & 0.029 & -0.020 \\
3 & 3 & 7 & 3 & 0.520 & 0.199 & 0.135 & 0.192 & 0.030 & -0.020 \\
3 & 4 & 7 & 3 & 0.531 & 0.200 & 0.134 & 0.195 & 0.030 & -0.019 \\
5 & 2 & 7 & 3 & 0.634 & 0.201 & 0.128 & 0.171 & 0.025 & -0.021 \\
5 & 3 & 7 & 3 & 0.632 & 0.202 & 0.128 & 0.177 & 0.027 & -0.019 \\
5 & 4 & 7 & 3 & 0.635 & 0.203 & 0.127 & 0.182 & 0.027 & -0.020 \\
7 & 2 & 7 & 3 & 0.754 & 0.202 & 0.119 & 0.155 & 0.026 & -0.010 \\
7 & 3 & 7 & 3 & 0.755 & 0.203 & 0.119 & 0.160 & 0.027 & -0.014 \\
7 & 4 & 7 & 3 & 0.754 & 0.204 & 0.118 & 0.167 & 0.025 & -0.021 \\
\\
5 & 3 & 5 & 2 & 0.740 & 0.268 & 0.181 & 0.172 & 0.022 & -0.024 \\
5 & 3 & 5 & 3 & 0.778 & 0.270 & 0.179 & 0.174 & 0.024 & -0.022 \\
5 & 3 & 5 & 4 & 0.811 & 0.269 & 0.173 & 0.175 & 0.026 & -0.021 \\
5 & 3 & 7 & 2 & 0.608 & 0.199 & 0.128 & 0.176 & 0.025 & -0.020 \\
5 & 3 & 7 & 3 & 0.632 & 0.202 & 0.128 & 0.177 & 0.027 & -0.019 \\
5 & 3 & 7 & 4 & 0.662 & 0.205 & 0.126 & 0.178 & 0.028 & -0.019 \\
5 & 3 & 9 & 2 & 0.545 & 0.161 & 0.099 & 0.179 & 0.030 & -0.015 \\
5 & 3 & 9 & 3 & 0.567 & 0.163 & 0.099 & 0.182 & 0.029 & -0.016 \\
5 & 3 & 9 & 4 & 0.583 & 0.167 & 0.098 & 0.181 & 0.030 & -0.019 \\
\\
\multicolumn{4}{r|}{mean} & 0.648 & 0.206 & 0.131 & 0.177 & 0.027 & -0.019 \\
\multicolumn{4}{r|}{$\sigma_\mathrm{sys}$} & 0.096 & 0.032 & 0.025 & 0.010 & 0.002 & 0.003 \\
\multicolumn{4}{r|}{$\sigma_\mathrm{stat}$} & 0.033 & 0.007 & 0.004 & 0.008 & 0.004 & 0.003 \\
\multicolumn{4}{r|}{$\sigma_\mathrm{sys}$/$\sigma_\mathrm{stat}$} & 3.0 & 4.9 & 6.3 & 1.2 & 0.7 & 1.2 \\
\bottomrule
\end{tabular}
\end{table}

% --------------------
\subsection{Using hospitalization data to characterize growth}
% --------------------

Public data from regions with significant CoViD-19 hospitalizations can
be used as an alternative measure of the exponential growth rate parameter $\delta$.
In the simplified homogeneous model assumed for this paper,
daily hospital admissions and the number of individuals in hospital
will also follow exponential growth or decline during periods of
constant transmission rate.
For the Ontario public hospitalization data corresponding to the case data
used for the fits in Table~\ref{tab:ont_fits}, there is sufficient data
to estimate $\delta_2$. 
The results are found to be
$\hat\delta_{2h} = -0.003 \pm 0.012$ (in-hospital data) and 
$\hat\delta_{2i} = -0.022 \pm 0.010$ (in-ICU data),
both consistent with the case data estimate ($\hat\delta_2=-0.019\pm0.003$), albeit with
much larger statistical uncertainties, given the limited number of hospitalizations
as compared to positive tests.

Differences between case data and hospitalization estimates 
for $\delta$ could arise in models with
heterogeneous populations, accounting for the fact that
the population requiring hospitalization for CoViD-19 is older
than the population tested positive for the disease,
and the transmission rates and circulation periods may differ by age.
In British Columbia, for example, the median age for the hospitalized population is
69 years, compared to 51 years for the population who have tested positive~\cite{BCCDC}.

\newpage

% --------------------
\section{Analyses of German state data}
% --------------------

Data from German states represent samples from populations subject to the same
public health measures and testing policies.
Germany enacted strict lock-down measures on March 22, 2020 as cases were rising rapidly.
Following the decline in cases in April, the measures were relaxed on May 6, 2020.
The Robert Koch Institute\cite{Koch} 
reports daily CoViD-19 cases and deaths for each of the 16 German states.
This section reports on analyses of the 13 states reporting more than 2000 cases by the end of June 2020.
Data between March 1 and June 25 are used, with exceptions that data after May 30 for Berlin and
after June 18 for North Rhine-Westphalia are not used since localized outbreaks affected the case
data for those periods.

\begin{figure}[ht]
    \centering
        \includegraphics[scale=.82, trim= 30 60 40 80, clip=true]{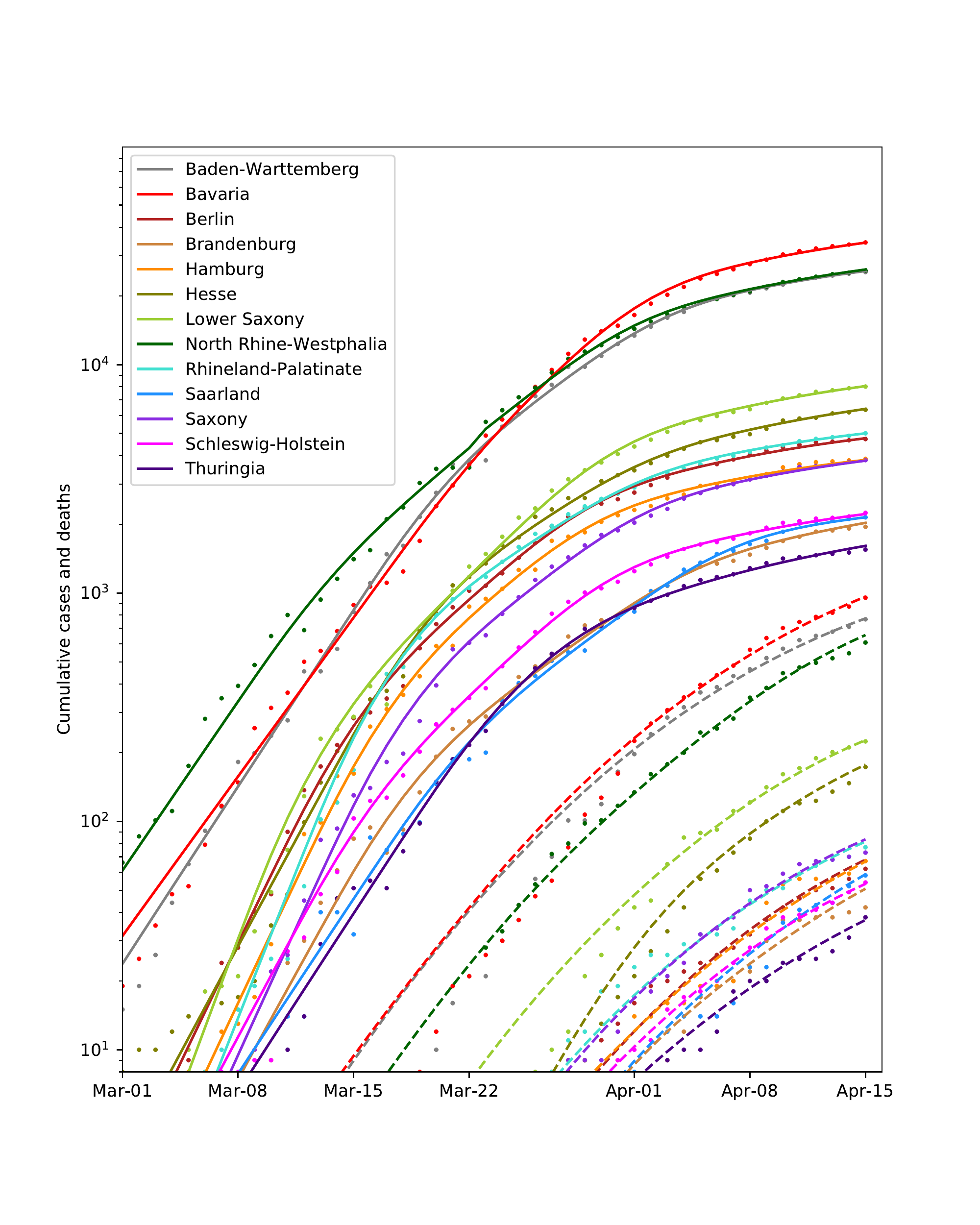}
    \caption{Cumulative case and death data (points) are shown in comparison to model predictions (lines) 
    with the point estimates shown in Table~\ref{tab:germany_point} for the period March 1 - April 15, 2020.
    The dashed lines show the model predictions for the cumulative deaths during this period.}
    \label{fig:germany_early}
\end{figure}

A minimum of three constant growth periods are necessary to describe the observed rise and fall in daily cases
in each of the states, with one transition (on day $d_1$) occurring before March 22 and another ($d_2$) near March 22.
A third transition is forced on May 6, in order to estimate the growth following relaxation.
Results from model fits to the data, showing the point estimates for the
transition dates $d_1$ and $d_2$ and the growth rates
are shown in Table~\ref{tab:germany_point}.
Figure~\ref{fig:germany_early} shows the cumulative distributions for cases and deaths for the first 45 days
compared to the model predictions that use the fitted parameter values.

\begin{table}[b]
\small\sf\centering
\caption{Point estimates for the four growth parameters from model fits to case data are shown for each state in Germany.
The transition day estimates $\hat d_1$ and $\hat d_2$ were determined in the fit, and are shown as the day in March.
Transition day $d_3$ was fixed to May 6, in order to estimate the growth parameter, $\delta_3$ following relaxation
of lock-down measures.}
\label{tab:germany_point}
\begin{tabular}{l c r c r c c}
\toprule
State & $\hat\delta_0$ & $\hat d_1$ & $\hat\delta_1$ & $\hat d_2$ & $\hat\delta_2$ & $\hat\delta_3$ \\
\midrule
Baden-Warttemberg & 0.288 & 12 & 0.105 & 24 & -0.049 & -0.060\\
Bavaria & 0.256 & 14 & 0.160 & 23 & -0.049 & -0.041\\
Berlin & 0.402 & 7 & 0.131 & 20 & -0.038 & -0.021\\
Brandenburg & 0.345 & 10 & 0.102 & 26 & -0.031 & -0.086\\
Hamburg & 0.410 & 9 & 0.122 & 20 & -0.043 & -0.061\\
Hesse & 0.354 & 11 & 0.078 & 23 & -0.031 & -0.026\\
Lower Saxony & 0.494 & 5 & 0.160 & 21 & -0.039 & 0.002\\
North Rhine-Westphalia & 0.274 & 5 & 0.134 & 22 & -0.038 & -0.029\\
Rhineland-Palatinate & 0.483 & 9 & 0.075 & 23 & -0.047 & -0.028\\
Saarland & 0.282 & 13 & 0.108 & 28 & -0.073 & -0.031\\
Saxony & 0.426 & 10 & 0.108 & 23 & -0.045 & -0.033\\
Schleswig-Holstein & 0.355 & 8 & 0.148 & 21 & -0.041 & -0.056\\
Thuringia & 0.287 & 15 & 0.117 & 20 & -0.017 & -0.029\\
\bottomrule
\end{tabular}
\end{table}

%left bot right top

\begin{figure}[t]
    \centering
        \includegraphics[scale=.68, trim= 5 20 40 40, clip=true]{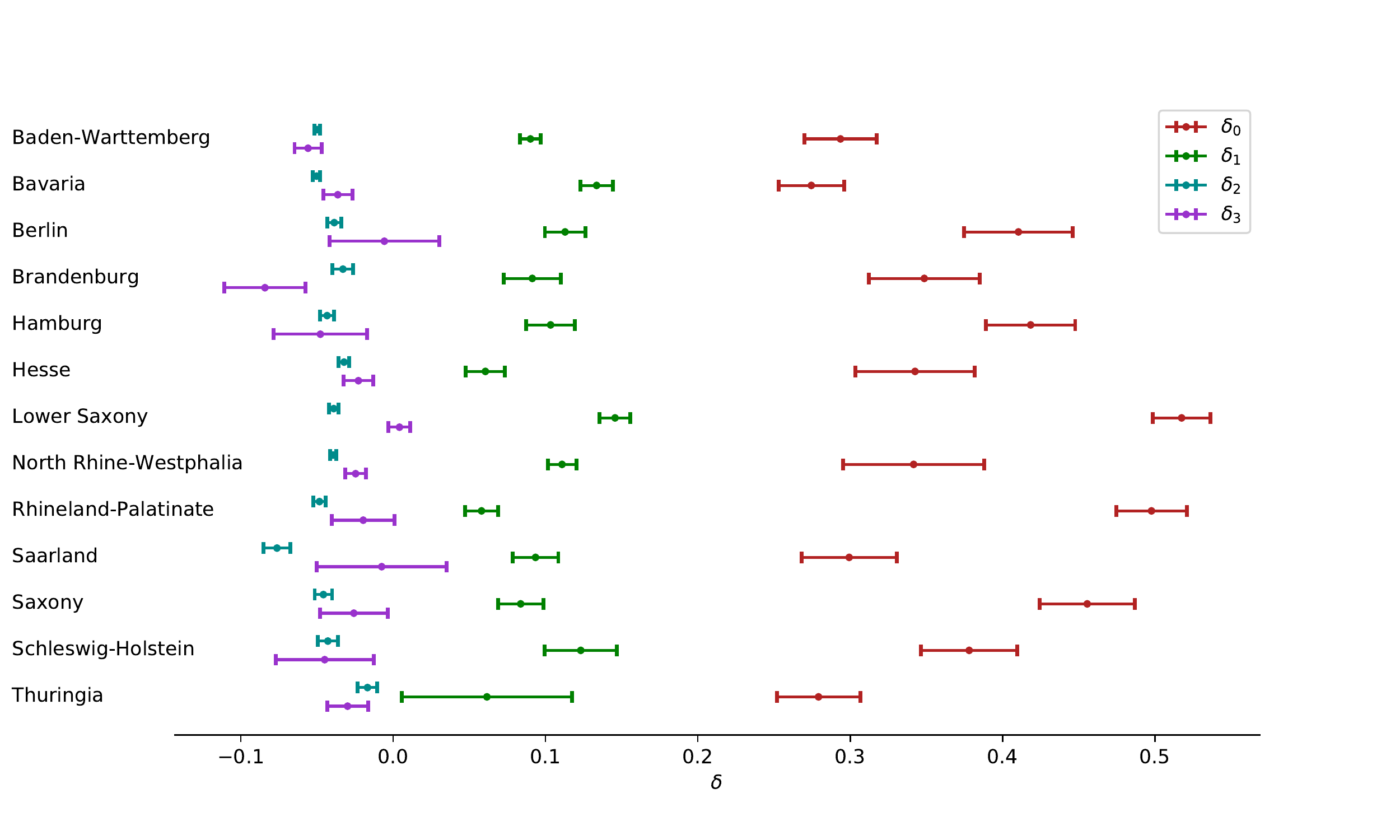}
    \caption{Summary of the growth rate estimates (corrected for bias) and their one standard deviation
    statistical uncertainties for the four periods for each state in Germany.}
    \label{fig:germany_deltas}
\end{figure}

To estimate bias and standard deviations of the estimators, 200 simulated samples were produced using
these point estimates, and fit keeping the transition dates fixed.
The mean and standard deviations of the estimates are reported in Table~\ref{tab:germany_sims}.
In separate studies, the standard deviations of transition date estimates were found to be typically 1 day.
The estimated dates for the second transition has a mean of 22.6 days following March 1, and a standard deviation 2.4 days. 
The growth estimates, corrected for bias, are summarized along with their estimated standard deviations
in Fig.~\ref{fig:germany_deltas}.

\begin{table}[t]
\small\sf\centering
\caption{To estimate bias and the standard deviations of estimators for the growth parameters, fits to 200 simulated
samples using the point estimates from Table~\ref{tab:germany_point} were performed. The means and standard deviations
of the estimates, shown here, do not indicate significant bias.} 
\label{tab:germany_sims}
\begin{tabular}{l c c c c}
\toprule
State & $\delta_0\ (\mu,\sigma)$ & $\delta_1\ (\mu,\sigma)$  & $\delta_2\ (\mu,\sigma)$  & $\delta_3\ (\mu,\sigma)$  \\
\midrule
Baden-Warttemberg & 0.283, 0.024 & 0.119, 0.007 & -0.048, 0.002 & -0.065, 0.009\\
Bavaria & 0.238, 0.022 & 0.186, 0.011 & -0.049, 0.002 & -0.046, 0.010\\
Berlin & 0.394, 0.036 & 0.150, 0.013 & -0.038, 0.005 & -0.036, 0.036\\
Brandenburg & 0.342, 0.036 & 0.113, 0.019 & -0.029, 0.007 & -0.088, 0.027\\
Hamburg & 0.401, 0.029 & 0.141, 0.016 & -0.043, 0.005 & -0.075, 0.031\\
Hesse & 0.365, 0.039 & 0.096, 0.013 & -0.031, 0.004 & -0.029, 0.010\\
Lower Saxony & 0.470, 0.019 & 0.175, 0.010 & -0.038, 0.003 & -0.001, 0.007\\
North Rhine-Westphalia & 0.207, 0.046 & 0.156, 0.009 & -0.037, 0.002 & -0.033, 0.007\\
Rhineland-Palatinate & 0.467, 0.023 & 0.092, 0.011 & -0.046, 0.004 & -0.037, 0.021\\
Saarland & 0.264, 0.031 & 0.122, 0.015 & -0.069, 0.009 & -0.055, 0.043\\
Saxony & 0.396, 0.031 & 0.132, 0.015 & -0.045, 0.006 & -0.041, 0.022\\
Schleswig-Holstein & 0.332, 0.032 & 0.173, 0.024 & -0.040, 0.007 & -0.068, 0.032\\
Thuringia & 0.295, 0.027 & 0.173, 0.056 & -0.016, 0.006 & -0.029, 0.013\\
\bottomrule
\end{tabular}
\end{table}

\clearpage

% --------------------
\subsection{Test of statistical uncertainty of growth rate estimates}\label{sec:test_stat}
% --------------------

\begin{figure}[t]
    \centering
        \includegraphics[scale=.66, trim= 0 40 20 80, clip=true]{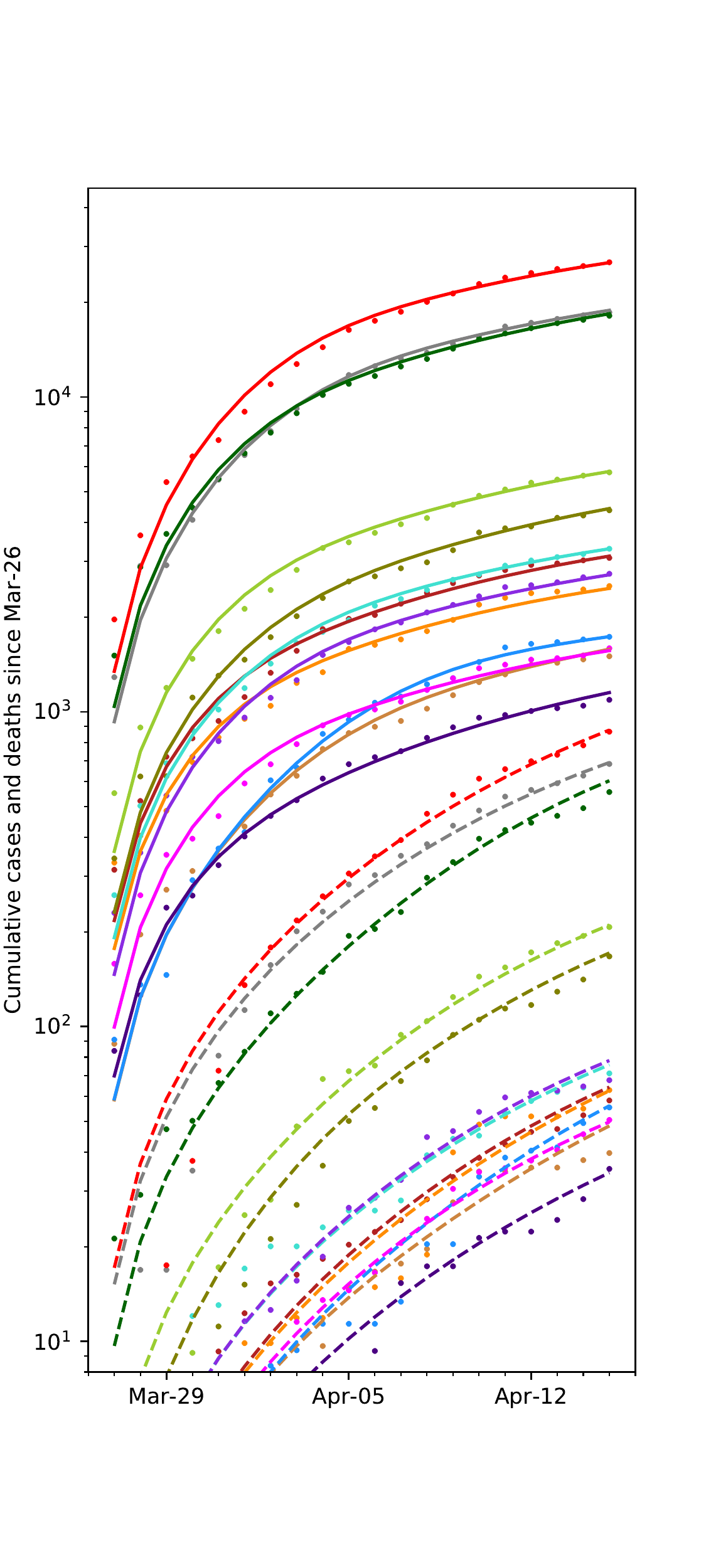}
        \includegraphics[scale=.66, trim= 0 40 20 80, clip=true]{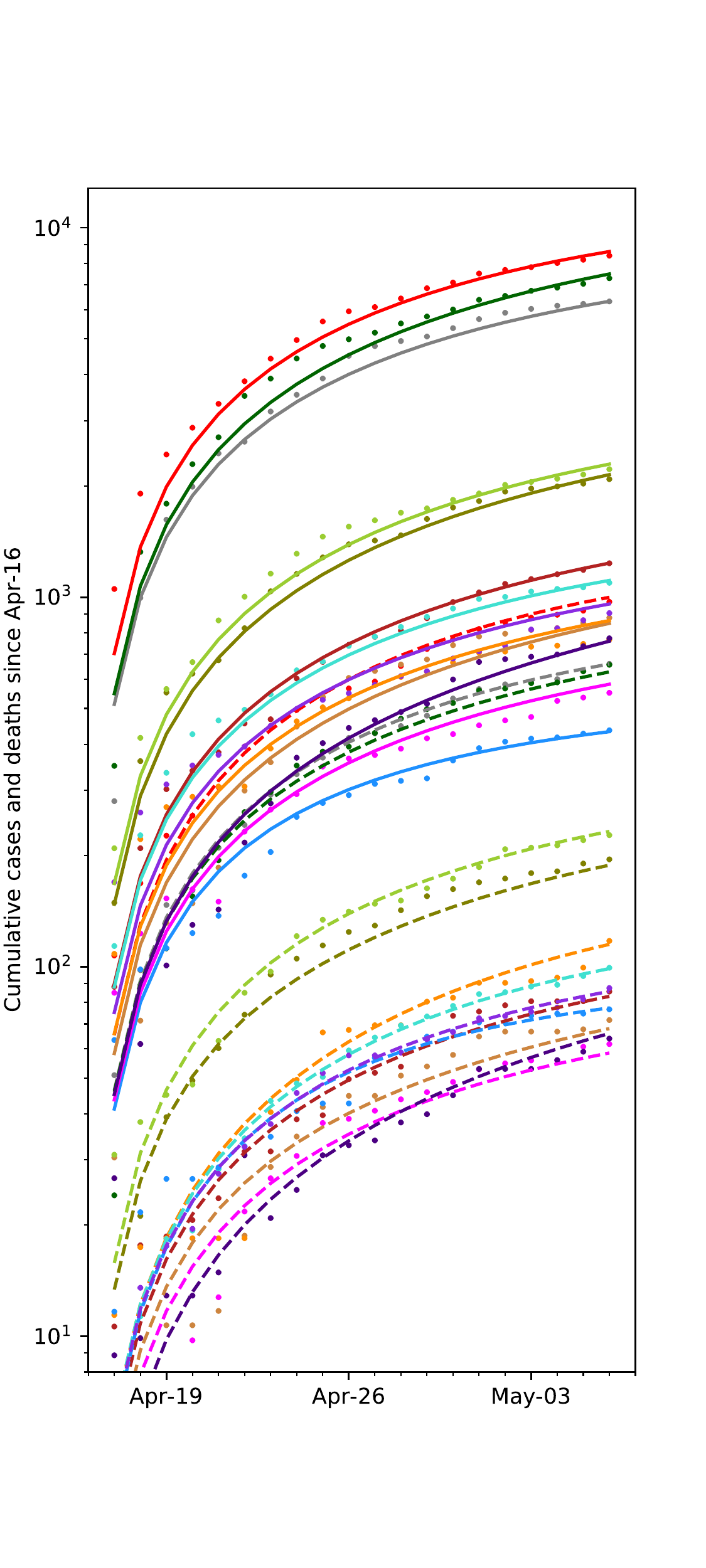}
    \caption{Cumulative case and death data (points) are shown in comparison to model predictions (lines) 
    with the point estimates shown in Table~\ref{tab:germany_point} for the two periods (A-left) March 27 - April 15
    and (B-right) April 17 - May 6. Lock-down measures were fixed across these two periods.
    The dashed lines show the model predictions for the cumulative deaths. The colors in the legend for
    Fig.~\ref{fig:germany_early} apply.}
    \label{fig:germany_AB}
\end{figure}

A test was performed for the statistical uncertainties assigned for growth rate estimates, by
comparing estimates for the growth rates for two independent periods (A) March 27 - April 15 and (B) April 17 - May 6.
Lock-down measures were fixed during this period, so growth rates would be expected to be nearly the same for
the two periods.
The cumulative cases and deaths for those periods are shown in Fig.~\ref{fig:germany_AB}

\begin{figure}[t]
    \centering
        \includegraphics[scale=.6]{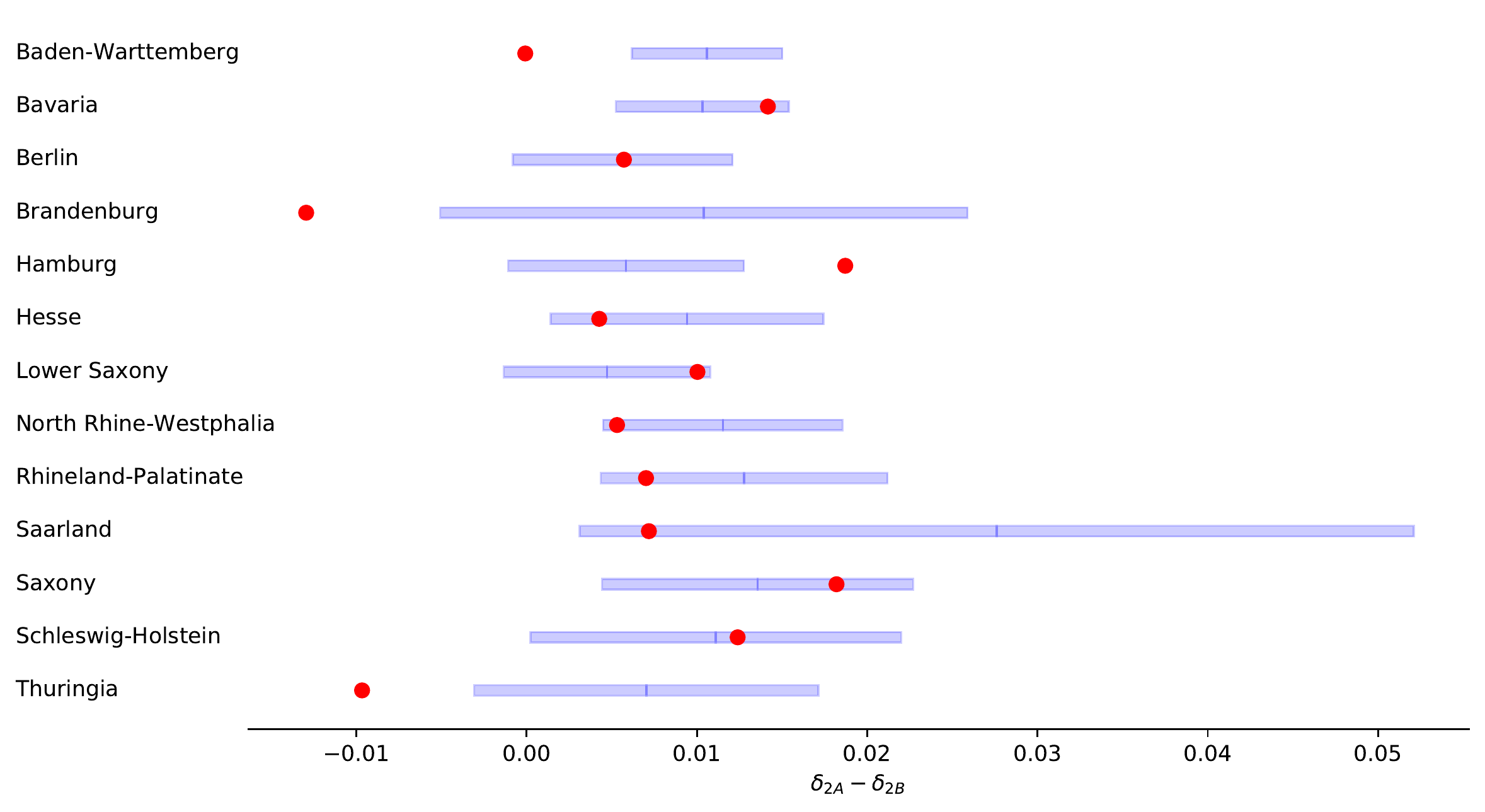}
    \caption{Difference in growth rate estimates for periods (A) and (B) during which lock-down
    measures were fixed. The red points show the data estimates. The blue bars show the mean and
    $\pm$ 1 standard deviations of the estimates from 100 simulated samples. The data estimates are
    consistent with being outcomes of the estimators defined by model fits: the goodness of fit test
    yields the p-value$=0.15$.}
    \label{fig:germany_delta2AB}
\end{figure}

Model fits were performed using case data for those periods separately, by optimizing only the
growth rate parameter, $\delta_2$, and the scale parameter $C_0$ (the initial size of
the contagious population).
The same approach was applied to 100 simulated samples using the point estimates shown in 
Table~\ref{tab:germany_point}.
The differences between the point estimates for the two periods, $\Delta = \hat\delta_{2A}-\hat\delta_{2B}$, 
for data and the simulated samples are summarized in Fig.~\ref{fig:germany_delta2AB}.
The hypothesis that the model correctly assigns a statistical uncertainty is tested using
the $\chi^2$ statistic, using the differences between the observed and expected $\Delta$ and the
model estimates for $\sigma_\Delta$.
The statistic yields $\chi^2=18.1$ for 13 degrees of freedom, with the corresponding p-value$ = 0.15$.

% --------------------
\subsection{Change in growth rate following relaxation of lock-down measures}
% --------------------

As described above, a transition in transmission rates was forced on May 6, so that the estimated growth rate parameters
before ($\hat\delta_2$) and after ($\hat\delta_3$) that date can be compared.
The test of bias and statistical uncertainty shown in section~\ref{sec:test_stat}
confirms the method to estimate the statistical significance of the observed difference in growth rate.
After correcting the point estimates in Table~\ref{tab:germany_point} for biases deduced from 
fits to simulated samples as reported in Table~\ref{tab:germany_sims},
the weighted average difference of the point estimates for the 13 German states 
is $\hat\delta_3 -\hat\delta_2 = 0.0145\pm0.0035$, a small but statistically significant increase.
Even with 50 days of data following relaxation, the growth rate estimates $\hat\delta_3$ have large uncertainty
due to the diminishing numbers of cases.

% --------------------
\subsection{Estimated size of epidemic}
% --------------------

\begin{figure}[t]
    \centering
        \includegraphics[scale=.6]{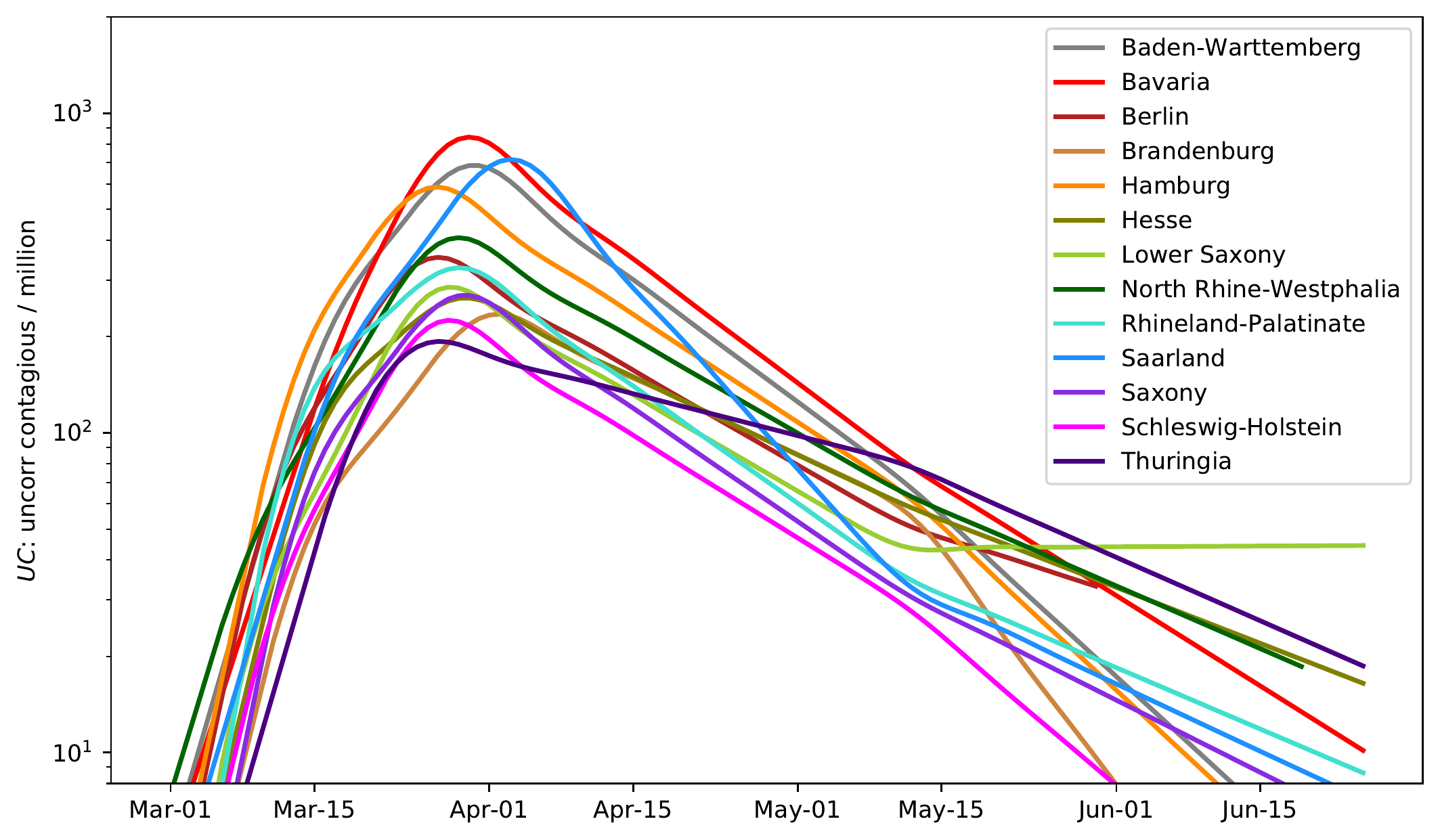}
    \caption{The uncorrected circulating contagious population history is shown for each state relative to
    the state population.}
    \label{fig:germany_UC}
\end{figure}

As discussed in section~\ref{sec:uc}
the ``uncorrected circulating contagious population'' (or $UC$) is a useful comparative statistic
for the size of the contagious population, with reduced systematic uncertainty.
The $UC$ history for each of the German states is shown in Fig.~\ref{fig:germany_UC} relative to the state population.
The relative sizes for different states were within a factor of about 5 throughout the period shown.

% --------------------
\subsection{Estimated death delay time distribution}
% --------------------

\begin{figure}[t]
    \centering
        \includegraphics[scale=.6]{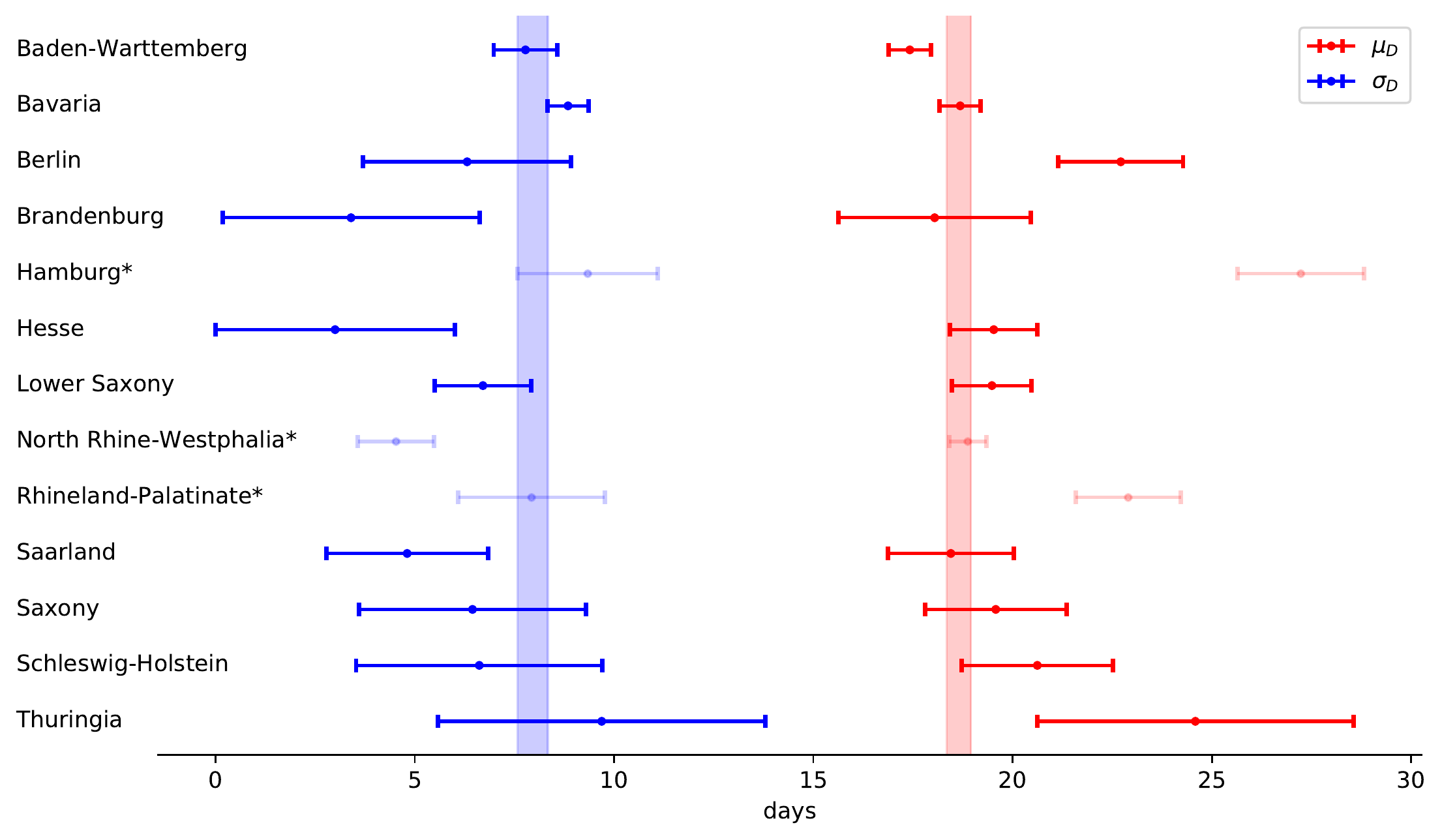}
    \caption{Estimates of the mean ($\mu_D$) and standard deviation ($\sigma_D$) of the time between
    becoming contagious and death. The states indicated with '*' and shaded out have one of their estimates
    more than $3\sigma$ from the full weighted average and are not used in the reported weighted averages.
    The vertical bars indicate the weighted averages and 68\% CL statistical uncertainty.}
    \label{fig:germany_death}
\end{figure}

After estimating the transmission rates using the case data, the death delay distribution parameters are estimated
in separate fits of the model to the cumulative death data.
In these fits, the transmission rates are fixed to the point estimates in Table~\ref{tab:germany_point} and only
three parameters are estimated; a scale factor (the fraction of the contagious who die), and the mean and standard
deviation of the time between becoming contagious and death.
The statistical uncertainty in these estimates are defined by the standard deviation of
estimates from fits to 100 simulated samples.
The results from each state are shown in Fig.~\ref{fig:germany_death}.
The data from the rapid rise in the death rate in the early stages of the epidemic carry the most information to
estimate the standard deviation of the delay ($\sigma_D$).
With larger values for $\sigma_D$, the rise in deaths is slower.
The analysis is not sensitive to a long upper tail in the delay distribution, and the assumption that the
distribution is normal may be incorrect.

The weighted average for the mean delay is $18.7\pm0.3$ days and for the standard deviation 
of the delay is $8.4\pm0.4$ days.
Sources of systematic uncertainty on the mean delay arise from possible mis-modelling of other delays.
Case and death report modelling have the latent delay in common, and since the fitted transition date $\hat d_2$, is
consistent with the expected transition (March 22), the systematic uncertainty from latent delay modelling 
could be assigned as the standard deviation of the observed transition dates, about 2.3 days.
Including an additional 2 days for uncertainty in case reporting delays, and using the mean 
latent period in the model (5 days), the time distribution from infection to death is estimated to
have a mean of $24 \pm 3$ days
and a standard deviation of about $8\pm1$ days.
The mean time from symptoms to death estimated with the pyPM model 2.3, $16.7$ is
consistent with the 95\% credible interval [16.9–19.2] reported from an early study of 
24 deaths in China.\cite{VERITY2020669}

% --------------------
\section{Summary and conclusions}
% --------------------

The simple pyPM model, introduced in this paper, can be used to
interpret public data in order to
characterize the spread of CoViD-19 in different regions around the world.
With relatively few parameters, 
the epidemic history can be conveniently summarized, with relatively long periods
having constant transmission rates (reported with the daily growth parameters $\delta_i$)
and the size of the contagious population (reported with $UC$).

As governments relax social distancing rules and open borders, 
it will be important to
compare the state of the disease and to detect changes in growth rates.
As the situation develops, results from additional studies
will be made available on the website www.pypm.ca.

%\begin{acks}
%This class file was developed by Sunrise Setting Ltd,
%Brixham, Devon, UK.\\
%Website: \url{http://www.sunrise-setting.co.uk}
%\end{acks}

\newpage

\bibliography{references}

\end{document}